\documentclass[10pt]{article} \usepackage[papersize={8.5in,11in}]{geometry} \usepackage{bm} \usepackage{graphicx} \usepackage{longtable}
\renewcommand{\vec}[1]{%
  \bm{\mathrm{#1}}%
}

\title{Tensor Manipulation in GPL Maxima}
\author{Viktor Toth\\
http://www.vttoth.com/}
\begin{document}
\maketitle
\begin{abstract}
GPL Maxima is an open-source computer algebra system based on DOE-MACSYMA. GPL Maxima included two tensor manipulation packages from DOE-MACSYMA, but these were in various states of disrepair. One of the two packages, CTENSOR, implemented component-based tensor manipulation; the other, ITENSOR, treated tensor symbols as opaque, manipulating them based on their index properties. The present paper describes the state in which these packages were found, the steps that were needed to make the packages fully functional again, and the new functionality that was implemented to make them more versatile. A third package, ATENSOR, was also implemented; fully compatible with the identically named package in the commercial version of MACSYMA, ATENSOR implements abstract tensor algebras.
\end{abstract}
\section{Introduction}
GPL Maxima (GPL stands for the GNU Public License, the most widely used open source license construct) is the descendant of one of the world's first comprehensive computer algebra systems (CAS), DOE-MACSYMA, developed by the United States Department of Energy in the 1960s and the 1970s. It is currently maintained by 18 volunteer developers, and can be obtained in source or object code form from {\tt http://maxima.sourceforge.net/}.
\par
Like other computer algebra systems, Maxima has tensor manipulation capability. This capability was developed in the late 1970s. Documentation is scarce regarding these packages' origins, but a select collection of e-mail messages by various authors survives, dating back to 1979-1982, when these packages were actively maintained at M.I.T.
\par
When this author first came across GPL Maxima, the tensor packages were effectively non-functional. The indicial tensor manipulation module could not display tensors, and could not carry out covariant differentiation; the component tensor manipulation package became nonresponsive the moment one tried to follow the documentation example and enter a metric.
\par
As it turned out, however, most of these problems were superficial in nature, and resulted from minor incompatibilities between various LISP systems. In other words, they were easy to correct. Much of the work invested into the tensor packages since that time involves new and improved functionality. The purpose of this paper is to document the work that has been done.
\section{Tensors in curved spacetime}
Let us begin with a brief review of the tensor algebra topics that are implemented by the tensor algebra packages in Maxima.
\subsection{Tensor transformation laws}
The mathematics of curved spacetime can be said to be based on the transformation properties of various quantities under a change of coordinates (see \cite{LR1989} for an excellent introduction.) When a switch is made to a new coordinate system (indicated by primed coordinates below), the new and old components of a vector relate to each other as follows:
\begin{equation}
\pmatrix{x'^1\cr\cr x'^2\cr\vdots\cr x'^n}=\pmatrix{\frac{\partial x'^1}{\partial x^1}&\frac{\partial x'^1}{\partial x^2}&\ldots&\frac{\partial x'^1}{\partial x^n}\cr\frac{\partial x'^2}{\partial x^1}&\frac{\partial x'^2}{\partial x^2}&\ldots&\frac{\partial x'^2}{\partial x^n}\cr\vdots&\vdots&\ddots&\vdots\cr\frac{\partial x'^n}{\partial x^1}&\frac{\partial x'^n}{\partial x^2}&\ldots&\frac{\partial x'^n}{\partial x^n}}\pmatrix{x^1\cr\cr x^2\cr\vdots\cr x^n}
\label{contra}
\end{equation}
Ordinary vectors that transform according to (\ref{contra}) are said to be {\it contravariant}. Loosely speaking, contravariant means ``just like a vector'', and {\it covariant} means ``just like a coordinate''. (By way of a simple example, consider a coordinate system that measures distances in meters. When you switch from meters to centimeters, your unit of measure decreases by a factor of 100. The quantities you measure, the length of a vector for instance, increase in contrast: a vector that was previously measured to be 5 units, i.e., 5 meters long, is now 500 units, i.e., 500 centimeters long. The coordinate units behave as covariant quantities, while the vector itself behaves as a contravariant quantity.)
\par
We can express the transformation laws for contravariant and covariant quantities in a more compact form. Denoting the transformation matrix in (\ref{contra}) by $\vec{A}$, we have:
\begin{equation}
\vec{x}'=\vec{A}\cdot\vec{x}
\end{equation}
\par
Now take another vectorial quantity, $\vec{y}$, that is known to produce an invariant inner product with $\vec{x}$:
\begin{equation}
\vec{x}\cdot\vec{y}=c
\end{equation}
How would this product change if we were to switch from one coordinate system to another? Why, it's supposed to be invariant, which means
\begin{equation}
\vec{x}'\cdot\vec{y}'=\vec{x}\cdot\vec{y}=c
\end{equation}
We know how to express $\vec{x}'$ as a function of $\vec{x}$ and the transformation matrix $\vec{A}$. But the same transformation could not apply to $\vec{y}$, because the result would not be the invariant quantity $c$:
\begin{equation}
\left(\vec{A}\cdot\vec{x}\right)\cdot\left(\vec{A}\cdot\vec{y}\right)\not=\vec{x}\cdot\vec{y}
\end{equation}
\par
Then again, if you think about this a moment, the expression $\vec{A}\cdot\vec{y}$ doesn't make much sense anyway. The vector $\vec{x}$ can be expressed in some coordinate system as a column vector. What do you multiply a column vector with in order to get a scalar? Why, a row vector of course. So $\vec{y}$ is a row vector. And how would you multiply a row vector on the left with the matrix $\vec{A}$? Well, you could do so in principle, but the result is most certainly not another row vector.
\par
You can, on the other hand, multiply a row vector on the right with a matrix. And perhaps, instead of using $\vec{A}$, you might consider using $\vec{A}^{-1}$, as in
\begin{equation}
\left(\vec{A}\cdot\vec{x}\right)\cdot\left(\vec{y}\cdot\vec{A}^{-1}\right)=\vec{x}\cdot\vec{y}
\end{equation}
If this equation is satisfied for a quantity $\vec{y}$, that quantity is said to be a covariant vector. A covariant vector maps a contravariant vector to the a scalar, namely the same scalar regardless of the coordinate system used.
\par
Many equations in physics and mathematics are matrix equations. For instance, we may have an equation in the form of
\begin{equation}
\vec{v}=\vec{T}\cdot\vec{u}
\end{equation}
When we change coordinate systems, we get:
\begin{equation}
\vec{v}'=\vec{A}\cdot\vec{v}=\vec{A}\cdot\vec{T}\cdot\vec{u}
\end{equation}
How would we express $\vec{v}'$ as a function of $\vec{u}'$? First, we multiply on the left by $\vec{T}^{-1}\vec{A}^{-1}$ (remembering that matrix multiplication is not commutative but it is associative):
\begin{equation}
\vec{T}^{-1}\cdot\vec{A}^{-1}\cdot\vec{v}'=\vec{T}^{-1}\cdot\vec{A}^{-1}\cdot\vec{A}\cdot\vec{T}\cdot\vec{u}=\vec{T}^{-1}\cdot\vec{T}\cdot\vec{u}=\vec{u}
\end{equation}
Next, we multiply on the left by $\vec{A}$:
\begin{equation}
\vec{A}\cdot\vec{T}^{-1}\cdot\vec{A}^{-1}\cdot\vec{v}'=\vec{A}\cdot\vec{u}=\vec{u}'
\end{equation}
Now that we managed to get $\vec{u}'$ on the right hand side, all that's left is to eliminate all factors of $\vec{v}'$ on the left by multiplying both sides with $\vec{A}\cdot\vec{T}\cdot\vec{A}^{-1}$:
\begin{equation}
\vec{A}\cdot\vec{T}\cdot\vec{A}^{-1}\cdot\vec{A}\cdot\vec{T}^{-1}\cdot\vec{A}^{-1}\cdot\vec{v}'=\vec{A}\cdot\vec{T}\cdot\vec{A}^{-1}\cdot\vec{u}'
\end{equation}
Once again remembering that matrix multiplication associative, we can reduce this equation to:
\begin{equation}
\vec{v}'=\left(\vec{A}\cdot\vec{T}\cdot\vec{A}^{-1}\right)\cdot\vec{u}'
\end{equation}
What we have here is a transformation rule for the matrix $\vec{T}$:
\begin{equation}
\vec{T}'=\vec{A}\cdot\vec{T}\cdot\vec{A}^{-1}
\end{equation}
We call a quantity like $\vec{T}$ a valence-{\tiny$\left[\matrix{1\cr 1}\right]$} tensor. It has 1 contravariant index, and 1 covariant index. Generally, a quantity that transforms as
\begin{equation}
\vec{X}'=\vec{A}^k\cdot\vec{X}\cdot\vec{A}^{-l}
\end{equation}
is called a valence-{\tiny$\left[\matrix{k\cr l}\right]$} tensor.
\par
Since the matrix $\vec{A}^{-1}$ appears in these equations, it is a good question how this matrix can be computed. Inverting $\vec{A}$ the ``hard way'' would be an obvious method, but there is an easier way. Notice that $\vec{A}^{-1}$ expresses a vector transformation just as $\vec{A}$ does:
\begin{equation}
\vec{x}=\vec{A}^{-1}\cdot\vec{A}\cdot\vec{x}=\vec{A}^{-1}\cdot\vec{x}'
\end{equation}
From this equation, we can read off the components of $\vec{A}^{-1}$ directly:
\begin{equation}
\overline{A}_{ij}=\frac{\partial x^i}{\partial x'^j}
\end{equation}

\subsection{Coordinate independence}
\par
Many tensor operations exist that result in values that are independent of the coordinate system in which the tensors are expressed. We have already seen one example: the product of a covariant and a contravariant vector. If we specify the coordinate system, this product can be expressed as:
\begin{equation}
\vec{x}\cdot\vec{y}=\sum_{i=1}^nx^iy_i
\end{equation}
where $n$ is the dimensionality of the vectors, and we use upper indices to distinguish contravariant vectors from covariant vectors that have lower indices.
\par
The important thing here is that this product does not depend on our choice of a coordinate system. The result will be the same in any coordinate system. This observation is the basis of the so-called {\it abstract index notation}: expressing tensor equations using indexed quantities without explicitly specifying the coordinate system.
\par
Another powerful notational convention is the Einstein summation convenction. Basically, it means dropping the summation sign in equations in which an index appears in both a contravariant and a covariant position; in these cases, the summation is implied, i.e.,
\begin{equation}
\sum_{i=1}^nx^iy_i=x^iy_i
\end{equation}
This operation, in which a repeated tensor index (often called a {\it dummy index}) disappears as an implied summation is carried out, is called {\it tensor contraction}.
\subsection{The metric}
The distance between two points in a rectilinear coordinate system can be expressed using the theorem of Pythagoras:
\begin{equation}
ds^2=\sum_i\left(dx^i\right)^2
\end{equation}
When a generalized coordinate system is used in curved space, this formula obviously does not work. It does, however, apply so long as the quantities $ds$ and $dx^i$ are infinitesimally small. How would the {\it infinitesimal squared distance}, $ds$, change under a change of coordinate systems? Here it is:
\begin{equation}
\left(ds'\right)^2=\sum_l\sum_k\sum_i\frac{\partial x'^i}{\partial x^k}\frac{\partial x'^i}{\partial x^l}dx^kdx^l
\end{equation}
By introducing the quantities
\begin{equation}
g_{kl}=\sum_i\frac{\partial x'^i}{\partial x^k}\frac{\partial x'^i}{\partial x^l}
\label{metric}
\end{equation}
we can rewrite the formula as
\begin{equation}
ds^2=g_{kl}dx^kdx^l
\end{equation}
Notice that
\begin{equation}
g_{kl}dx^l=\frac{\partial}{\partial x^k}g_{ml}dx^mdx^l=\frac{\partial}{\partial x^k}ds^2
\end{equation}
If we take the infinitesimal squared distance, $ds^2$, to be unity, multiplying by $g_{kl}$ accomplished a conversion from the infinitesimal quantity $dx^l$ to $\partial/\partial x^k$. This is called a {\it lowering of the index}. What we basically have here is a unique way, with the help of $g_{kl}$, to associate contravariant and covariant quantities.
\par
Similarly, the inverse of the metric tensor, $g^{kl}$, can be used the {\it raise} a tensor index. The inner product of the metric tensor with itself gives the Kronecker-delta (i.e., the identity matrix):
\begin{equation}
g_{kl}g^{lm}=\delta_k^m
\end{equation}
\par
The quantities $g_{kl}$ themselves transform as a valence-{\tiny$\left[\matrix{0\cr 2}\right]$} covariant tensor, which justifies the notation.
\subsection{Covariant differentiation}
In Euclidean space, when rectilinear coordinates are used, partial differentiation by a coordinate results in new tensorial quantities. For instance, partial differentiation of a scalar field by the coordinates defines the gradient of that field. Higher-order quantities can also be differentiated this way; we can, for instance, compute the gradient of a vector field and obtain a valence-{\tiny$\left[\matrix{1\cr 1}\right]$} tensor field.
\par
This is no longer true in curved space, where partial differentiation by a coordinate no longer necessarily produces a quantity that transforms as a tensor under a coordinate transformation. The question arises, then: is it possible to define a new type of a derivative operator, one that on the one hand has the same algebraic behavior (e.g., obeying the Leibnitz rule) as the ordinary derivative operator, but on the other hand, produces a quantity that behaves like a tensor under a change of coordinates?
\par
To answer this question, we first need to explore the concept of a parallel vector field. In flat space, a parallel vector field consists of identical vectors attached to each point in spacetime. In other words for a vector $\vec{X}$
\begin{equation}
dX^j=\sum_k\frac{\partial X^j}{\partial x^k}dx^k=0
\end{equation}
How does this expression transform under a change of coordinates? We can answer this question by utilizing the identity $X'^j=\sum_h\frac{\partial x'^j}{\partial x^h}X^h$, to get:
\begin{equation}
dX'^j=\sum_h\frac{\partial x'^j}{\partial x^h}dX^h+\sum_h\sum_k\frac{\partial^2x'^j}{\partial x^h\partial x^k}X^hdx^k
\label{parallel}
\end{equation}
With the help of the metric, it is possible to eliminate the second derivatives from this equation. First, we need to differentiate (\ref{metric}) with respect to $x^h$ and then cyclically permute the indices to obtain the following identities:
\begin{eqnarray}
&\frac{\partial g_{kl}}{\partial x^h}=\sum_j\left(\frac{\partial^2x'^j}{\partial x^h\partial x^k}\frac{\partial x'^j}{\partial x^l}+\frac{\partial x'^j}{\partial x^k}\frac{\partial^2x'^j}{\partial x^h\partial x^l}\right)\nonumber\\
&\frac{\partial g_{lh}}{\partial x^k}=\sum_j\left(\frac{\partial^2x'^j}{\partial x^k\partial x^l}\frac{\partial x'^j}{\partial x^h}+\frac{\partial x'^j}{\partial x^l}\frac{\partial^2x'^j}{\partial x^k\partial x^h}\right)\nonumber\\
&\frac{\partial g_{hk}}{\partial x^l}=\sum_j\left(\frac{\partial^2x'^j}{\partial x^l\partial x^h}\frac{\partial x'^j}{\partial x^k}+\frac{\partial x'^j}{\partial x^h}\frac{\partial^2x'^j}{\partial x^l\partial x^k}\right)
\end{eqnarray}
\par
Adding the first and second of these three equations, subtracting the third, and dividing the result by two, we obtain:
\begin{equation}
\Gamma_{hkl}=\frac{1}{2}\left(\frac{\partial g_{kl}}{\partial x^h}+\frac{\partial g_{lh}}{\partial x^k}-\frac{\partial g_{hk}}{\partial x^l}\right)=\sum_j\frac{\partial^2x'^j}{\partial x^h\partial x^k}\frac{\partial x'^j}{\partial x^l}
\end{equation}
Now we multiply (\ref{parallel}) by $\partial x'^j/\partial x^l$ and sum over $j$ to get
\begin{equation}
\sum_j\frac{\partial x'^j}{\partial x^l}dX'^j=\sum_hg_{hl}dX^h+\sum_h\sum_k\Gamma_{hkl}X^hdx^k
\end{equation}
Both sides of this equation should be zero for a parallel vector field. The right hand side can be made more meaningful if we multiply it by $g^{jl}$ (we're also dropping the summation signs now, relying on the summation convention instead):
\begin{equation}
dX^j+g^{jl}\Gamma_{hkl}X^hdx^k=0
\end{equation}
In other words, for a parallel vector field, the quantities $\Gamma_{hk}{}^j=g^{jl}\Gamma_{hkl}$ express the difference by which the ordinary differential operator will fail. This suggests a definition for a differential operator in curved space:
\begin{equation}
\nabla_kX^j=\frac{\partial X^j}{\partial x^k}+\Gamma_{hk}{}^jX^h
\end{equation}
For a covariant vector field, the same derivation yields
\begin{equation}
\nabla_kX_j=\frac{\partial X_j}{\partial x^k}-\Gamma_{jk}{}^hX_h
\end{equation}
And, for an arbitrary tensor of valence-{\tiny$\left[\matrix{r\cr s}\right]$}, we get
\begin{equation}
\nabla_kT^{j_1\ldots j_r}_{i_1\ldots i_s}=\frac{\partial T^{j_1\ldots j_r}_{i_1\ldots i_s}}{\partial x^k}+\sum_{\alpha=1}^r\Gamma_{mk}{}^{j_\alpha}T^{j_1\ldots j_{\alpha-1}mj_{\alpha+1}\ldots j_r}_{i_1\ldots i_s}-\sum_{\beta=1}^s\Gamma_{i_{\beta}k}{}^mT^{j_1\ldots j_r}_{i_1\ldots i_{\beta-1}mi_{\beta+1}\ldots i_s}
\end{equation}
\par
The quantities $\Gamma_{ijk}$ and $\Gamma_{ij}{}^k$ are called the Christoffel-symbols of the first and second kind, respectively.
\par
The operator we just defined, $\nabla_k$, can justifiably called a differential operator for several reasons. First, as can be verified by direct computation, it obeys the Leibnitz rule: $\nabla_k\left(\vec{X}\cdot\vec{Y}\right)=\left(\nabla_k\vec{X}\right)\cdot\vec{Y}+\vec{X}\cdot\left(\nabla_k\vec{Y}\right)$. Second, it can be shown that once the metric, $g_{ij}$, is given, the differential operator is uniquely defined, in that it is the only such operator that preserves the inner product of two vectors parallel transported along any curve, i.e., $\nabla_kg_{ij}v^iw^j=0$.
\par
Note that a shorthand is often used to express the covariant derivative; for instance, instead of $\nabla_kX_{ij}$, we often write $X_{ij;k}$.
\subsection{The Lie-derivative}
The covariant derivative is tensorial in nature, but unfortunately, it depends on the metric. The ordinary derivative is not dependent on the metric, but it's not tensorial: it depends on the choice of the coordinate system. So the question arises: is it possible to construct operators that are not dependent on the metric, but produce tensorial results that are not coordinate system dependent?
\par
There are, in fact, many such operators possible (see \cite{PR1990}), most notably among them the Lie-derivative:
\begin{equation}
\mathcal{L}_VH^{i\ldots k}_{l\ldots n}=V^h\partial_hH^{i\ldots k}_{l\ldots n}-H^{h\ldots k}_{l\ldots n}\partial_hV^i-\ldots-H^{i\ldots h}_{l\ldots n}\partial_hV^k+H^{i\ldots k}_{h\ldots n}\partial_lV^h+H^{i\ldots k}_{l\ldots h}\partial_nV^h
\end{equation}
The geometric meaning of the Lie-derivative is directional differentiation: it tells you how a tensor field changes in the direction of the vector field $\vec{V}$.
\subsection{Commutators}
The ordinary partial differential operator is known to commute: $\partial_i\partial_j\vec{X}-\partial_j\partial_i\vec{X}=0$. Is this also true for the covariant derivative?
\par
For a scalar field $f$, the commutator is
\begin{eqnarray}
&&\left(\nabla_k\nabla_l-\nabla_l\nabla_k\right)f=\nabla_k\frac{\partial f}{\partial x^l}-\nabla_l\frac{\partial f}{\partial x^k}=\nonumber\\
&&\frac{\partial^2f}{\partial x^k\partial x^l}-\Gamma_{lk}{}^h\frac{\partial f}{\partial x^h}-\frac{\partial^2f}{\partial x^k\partial x^l}+\Gamma_{kl}{}^h\frac{\partial f}{\partial x^h}=\left(\Gamma_{kl}{}^h-\Gamma_{lk}{}^h\right)\frac{\partial f}{\partial x^h}
\end{eqnarray}
The way we defined them, the Christoffel-symbols are symmetric in their first two indices, and therefore $\Gamma_{kl}{}^h-\Gamma_{lk}{}^h=0$. We could conceivably define another differential operator whose {\it connection coefficients} (the generalization of the concept embodied in the Christoffel-symbols) are not symmetrical, and thus $c_{kl}{}^h-c_{lk}{}^h$ does not vanish; this quantity is then called {\it torsion}.
\par
How about the commutator of the covariant derivative on a vector field? Here it is:
\begin{eqnarray}
&&\left(\nabla_k\nabla_l-\nabla_l\nabla_k\right)X^j=\nonumber\\
&&\frac{\partial}{\partial x^k}\left(\nabla_lX^j\right)+\Gamma_{mk}{}^j\left(\nabla_lX^m\right)-\Gamma_{lk}{}^m\left(\nabla_mX^j\right)-\frac{\partial}{\partial x^l}\left(\nabla_kX^j\right)-\Gamma_{ml}{}^j\left(\nabla_kX^m\right)+\Gamma_{kl}{}^m\left(\nabla_mX^j\right)=\nonumber\\
&&\left(\frac{\partial^2}{\partial x^k\partial x^l}-\frac{\partial^2}{\partial x^l\partial x^k}\right)X^j+\frac{\partial}{\partial x^k}\left(\Gamma_{hl}{}^jX^h\right)-\frac{\partial}{\partial x^l}\left(\Gamma_{hk}{}^jX^h\right)+\Gamma_{mk}{}^j\frac{\partial X^m}{\partial x^l}-\Gamma_{ml}{}^j\frac{\partial X^m}{\partial x^k}+\nonumber\\
&&\Gamma_{mk}{}^j\Gamma_{hl}{}^mX^h-\Gamma_{ml}{}^j\Gamma_{hk}{}^mX^h+\left(\Gamma_{kl}{}^m-\Gamma_{lk}{}^m\right)\left(\nabla_mX^j\right)=\nonumber\\
&&\frac{\partial\Gamma_{hl}{}^j}{\partial x^k}X^h+\Gamma_{hl}{}^j\frac{\partial X^h}{\partial x^k}-\frac{\partial\Gamma_{hk}^j}{\partial x^l}X^h-\Gamma_{hk}{}^j\frac{\partial X^h}{\partial x^l}+\Gamma_{mk}{}^j\frac{\partial X^m}{\partial x^l}-\Gamma_{ml}{}^j\frac{\partial X^m}{\partial x^k}+\nonumber\\
&&\Gamma_{mk}{}^j\Gamma_{hl}{}^mX^h-\Gamma_{ml}{}^j\Gamma_{hk}{}^mX^h+\left(\Gamma_{kl}{}^m-\Gamma_{lk}{}^m\right)\left(\nabla_mX^j\right)=\nonumber\\
&&\left(\frac{\partial\Gamma_{hl}{}^j}{\partial x^k}-\frac{\partial\Gamma_{hk}{}^j}{\partial x^l}+\Gamma_{mk}{}^j\Gamma_{hl}{}^m-\Gamma_{ml}{}^j\Gamma_{hk}{}^m\right)X^h+\left(\Gamma_{kl}{}^m-\Gamma_{lk}{}^m\right)\left(\nabla_mX^j\right)
\end{eqnarray}
\par
The second part of this result we already recognize: it's the torsion again. The first part contains the Riemann curvature tensor, defined as
\begin{equation}
R_{hlk}{}^j=-R_{hkl}{}^j=\frac{\partial\Gamma_{hl}{}^j}{\partial x^k}-\frac{\partial\Gamma_{hk}{}^j}{\partial x^l}+\Gamma_{mk}{}^j\Gamma_{hl}{}^m-\Gamma_{ml}{}^j\Gamma_{hk}{}^m
\end{equation}
Several other quantities are derived from the curvature tensor that are widely used in physics and mathematics. These include the Ricci-tensor:
\begin{equation}
R_{ij}=R_{ji}=R_{ijk}{}^k
\end{equation}
the scalar curvature:
\begin{equation}
R=R_i^i
\end{equation}
and the Weyl-tensor (\cite{W1984}):
\begin{equation}
W_{ijkl}=R_{ijkl}+\frac{2}{(n-1)(n-2)}Rg_{j[i}g_{l]k}+\frac{2}{n-2}\left(g_{k[i}R_{l]j}-g_{j[i}R_{l]k}\right)
\end{equation}
(Indices in square brackets indicate complete antisymmetrization; for instance, $A_{[ij]}=(A_{ij}-A_{ji})/2!$.)
\par
In gravitational theory, the Ricci-tensor expresses the contribution of matter and energy to the curvature tensor, while the Weyl tensor (often called the conformal tensor, and labelled $C_{abcd}$) represents the curvature of empty spacetime, e.g., curvature far from any matter and energy sources, caused by free gravitational waves.
\par
Yet another important tensor in gravitational theory is the Einstein tensor:
\begin{equation}
G_{ij}=R_{ij}-\frac{1}{2}Rg_{ij}
\end{equation}
\subsection{Torsion and nonmetricity}
As described in the previous section, if the connection coefficients are not symmetric in their indices, the metric is said to have torsion.
\par
Torsion can be characterized by the torsion tensor $\tau_{ij}{}^k$. When torsion is present, it modifies the connection coefficients such that:
\begin{equation}
c_{ijk}=\Gamma_{ijk}+\frac{1}{2}\left(\tau_{ij}{}^mg_{km}+\tau_{ki}{}^mg_{jm}+\tau_{kj}{}^mg_{im}\right)
\end{equation}
\par
The quantites used to modify the Christoffel symbols in this expression are often called {\it contortion coefficients}:
\begin{equation}
\kappa_{ijk}=-\frac{1}{2}\left(\tau_{ij}{}^mg_{km}+\tau_{ki}{}^mg_{jm}+\tau_{kj}{}^mg_{im}\right)
\end{equation}
\par
Using the derivative operators based on the Christoffel-symbols (with or without torsion) we find that the covariant derivative of the metric tensor is identically zero:
\begin{equation}
\nabla_kg_{ij}=0
\end{equation}
It is possible to modify the derivative operator, while retaining its algebraic properties, so that the covariant derivative of the metric tensor is no longer zero. This property of a derivative operator is called nonmetricity. Nonmetricity is characterized by the vector field $\mu_k$:
\begin{equation}
g_{ij;k}=-\mu_kg_{ij}
\end{equation}
The contribution of the nonmetricity field to the connection coefficients can be expressed as
\begin{equation}
c_{ijk}=\Gamma_{ijk}+\frac{1}{2}\left(g_{ik}\mu_j+g_{jk}\mu_i-g_{ij}\mu_k\right)
\end{equation}
The quantities that appear in this equation are sometimes called {\it nonmetricity coefficients}:
\begin{equation}
\nu_{ijk}=\frac{1}{2}\left(-g_{ik}\mu_j-g_{jk}\mu_i+g_{ij}\mu_k\right)
\end{equation}
\subsection{Rigid frames}
In a smooth space, the immediate vicinity of every point can be approximated by a flat space (i.e., a space without curvature.) This gives rise to the idea of attaching a rigid (typically, though not necessarily, rectilinear) set of basis vectors (a {\it rigid frame}, also called (in four dimensions) a tetrad, {\it vierbein}, or, in arbitrary dimensions, a {\it vielbein}) to each point. While this can be done at each point, the resulting collection of basis vectors will not form a coordinate system. Intuitively, this means that supposedly parallel lines in a coordinate grid will, sooner or later, meet and cross each other. Such a basis is said to be {\it nonholonomic}.
\par
Though not a coordinate base, such a base of {\it moving frames} can nevertheless be used to measure vectors and other quantities in that space. And, at least in some cases, the use of a nonholonomic frame can result in simpler equations.
\par
A nonholonomic base is characterized by a set of $n$ basis vectors $e_{(a)}$, where $n$ is the dimensionality of the space, and parenthesized indices refer to the basis vectors themselves. Each of these basis vectors can have (contravariant) components relative to some coordinate base: $e_{(a)}^i$. The inner product of basis vectors defines the {\it frame metric}: $e_{(a)}^ie_{(b)i}=\eta_{ab}$. Typically, the frame metric would either be the unit matrix (for a Euclidean frame) or a unit matrix with Minkowski signature (for a Lorentz frame.)
\par
Raising and lowering of the frame basis indices can be accomplished using the frame metric or its inverse; e.g., $e^{(a)}=\eta^{ab}e_{(b)}$. Since $\eta_{ab}\eta^{bc}=\delta_a^c$, it's also true that $e_i^{(a)}e^k_{(a)}=\delta_i^k$.
\par
To compute the covariant derivative of a quantity expressed using these basis vectors, we need the {\it Ricci rotation coefficients} (\cite{LL1975}):
\begin{equation}
\gamma_{abc}=e_{(a)i;k}e_{(b)}^ie_{(c)}^k
\end{equation}
In a frame base, these coefficients take the place of the Christoffel symbols. Computing them in this form is not practical, however, since we'd need a way to compute the covariant derivative of the basis vectors before the coefficients are calculated. What we do instead is take a linear combination of these coefficients:
\begin{equation}
\lambda_{abc}=\gamma_{abc}-\gamma_{acb}=\left(e_{(a)i;k}-e_{(a)k;i}\right)e_{(b)}^ie_{(c)}^k
\end{equation}
The quantities $\lambda_{abc}$ or $\lambda_{ab}{}^c$ are often called the {\it frame bracket}.
\par
The last expression can be further simplified, with ordinary partial differentiation replacing the covariant derivative, as the Christoffel symbols cancel out due to their index symmetries, leaving only the torsion:
\begin{equation}
\left(e_{(a)i;k}-e_{(a)k;i}\right)e_{(b)}^ie_{(c)}^k=\left(e_{(a)i,k}-e_{(a)k,i}-\tau_{ik}{}^me_{(a)m}\right)e_{(b)}^ie_{(c)}^k
\end{equation}
Using these quantities, the Ricci rotation coefficients can be expressed as
\begin{equation}
\gamma_{abc}=\frac{1}{2}\left(\lambda_{abc}+\lambda_{bca}-\lambda_{cab}\right)
\end{equation}
The Ricci rotation coefficients are antisymmetric in their first two indices: $\gamma_{abc}=-\gamma_{bac}$. This means that in contrast with the $n^2(n+1)/2$ independent components of the Christoffel-symbols, the Ricci rotation coefficients have only $n^2(n-1)/2$ independent components: 24 instead of 40 in the case of 4 dimensions. This explains why sometimes, the use of a frame base presents significant advantages.
\par
Expressed using the Ricci rotation coefficients, the Riemann tensor takes the following form (cf. \cite{W1984}, eq. 3.4.21):
\begin{equation}
R_{dabc}=e_{(a)}^i\nabla_{i}\gamma_{bcd}-e_{(b)}^i\nabla_{i}\gamma_{acd}+\eta^{ef}\left(\gamma_{afc}\gamma_{bed}-\gamma_{bfc}\gamma_{aed}+\gamma_{afb}\gamma_{ecd}-\gamma_{bfa}\gamma_{ecd}\right)
\end{equation}
\subsection{Algebraic classification of metrics}
When studying curved manifolds, one frequently encountered problem is whether or not two metrics represent the same manifold. Usually it is not at all obvious from the algebraic properties alone if two metrics are equivalent; consider, for instance, the metric associated with polar vs. rectangular coordinates ($dx^2+dy^2$ vs. $dr^2+r^2d\phi^2$) or, in 3 dimensions, Cartesian vs. spherical coordinates ($dx^2+dy^2+dz^2$ vs. $dr^2+r^2d\theta^2+r^2\sin^2{\theta}d\phi^2$). A variety of techniques exists to help establish if two metrics are guaranteed to be different or may be the same; most well known among these, perhaps, is the Petrov-classification, applicable to 4-dimensional metrics with Minkowski signature.
\par
The Petrov-classification is based on the fact that the Weyl-tensor of any metric gives rise to several invariant quantities. Denoted by the symbols $\Psi_0, \Psi_1\ldots\Psi_4$, these invariants provide the coefficients of a quartic equation for the principal null directions of the metric (see \cite{SKMHH2003} for details.)
\par
To compute the Petrov classification, one requires two elements: the Weyl tensor in a coordinate base, and a Newman-Penrose null tetrad representing the metric. For instance, if we have an orthonormal tetrad base $e_{(a)}$ with $(+,-,-,-)$ signature and coordinates $(t,x,y,z)$, the corresponding null tetrad can be computed as:
\begin{eqnarray}
&k=\frac{\sqrt{2}}{2}\left(e_1+e_2\right)\nonumber\\
&l=\frac{\sqrt{2}}{2}\left(e_1-e_2\right)\nonumber\\
&\overline{m}=\frac{\sqrt{2}}{2}\left(e_3+ie_4\right)\nonumber\\
&m=\frac{\sqrt{2}}{2}\left(e_3-ie_4\right)
\end{eqnarray}
With the Weyl-invariants and the null tetrad at hand, classification of the metric can proceed. Two principal quantities used in this computation are
\begin{equation}
I=\Psi_0\Psi_4-4\Psi_1\Psi_3+3\Psi_2^2
\end{equation}
and
\begin{equation}
J=\left|\matrix{\Psi_0&\Psi_1&\Psi_2\cr\Psi_1&\Psi_2&\Psi_3\cr\Psi_2&\Psi_3&\Psi_4}\right|
\end{equation}
The actual algorithm is quite messy: the Petrov type is determined by which of the $\Psi$ invariants are non-zero; in some cases, this pattern directly determines the Petrov type, while in other cases, $I$, $J$, and additional coefficients need to be computed first. A complete description of a modern algorithm can be found in (\cite{PSD2000}).
\par
The algebraic types obtained by the Petrov-classification can be represented in a diagrammatic form illustrating increased specialization:
$$
\matrix{I\cr
\downarrow&\searrow\cr
II&\rightarrow&D\cr
\downarrow&\searrow&\downarrow&\searrow\cr
III&\rightarrow&N&\rightarrow&0}
$$
\par
These algebraic types also possess geometric significance. They determine the number of {\it principal null directions} that a metric possesses. A general four-dimensional spacetime has four principal null directions, but in the algebraically special cases, some of these null directions coincide:
\begin{center}
\begin{tabular}{ccl}
\\
Type&&Null directions\\
\hline\\[-6pt]
I&$(1,1,1,1)$&All null directions are independent\\
D&$(2,2)$&Two pairs of coinciding null directions\\
II&$(2,1,1)$&Two null directions coincide, two are independent\\
III&$(3,1)$&One triplet of coinciding null directions, one independent\\
N&$(4)$&All null directions coincide\\
\\
\end{tabular}
\end{center}
\par
For a more detailed explanation, see \cite{PR1990} and \cite{SKMHH2003}.
\subsection{Exterior forms}
A particularly interesting variety of tensors are fully antisymmetric covariant tensors. I.e., tensors of the type $A_{i\ldots k\ldots l\ldots n}=-A_{i\ldots l\ldots k\ldots n}$.
\par
These tensors are notable because they form the basis of Cartan's {\it exterior calculus of differential forms}. Simple operations exist on these tensors that can be expressed in an index-free notation. They include the exterior derivative, contraction with a vector, and the exterior, or wedge product.
\par
The wedge product is the fully antisymmetrized product of two covariant, fully antisymmetric tensors. Unfortunately, not all authors agree on the precise definition of the wedge product. As Penrose and Rindler comment in \cite{PR1990}: ``We should remark that, in the literature, a slightly different convention is frequently employed, in that the quantities $a_{\alpha_1\ldots\alpha_p}:=p!A_{\alpha_1\ldots\alpha_p}$ rather than our $A_{\alpha_1\ldots\alpha_p}$ are used to denote the components of a $p$-form."
\par
The advantage of the notation employed by Penrose and Rindler, and also used in \cite{JAW1984} and \cite{LR1989}, is that it agrees with the concept of total antisymmetrization. The other notation, used for instance in \cite{GS1989} or \cite{F2004}, is more geometrically inspired, as it produces exterior forms that correspond with the notion of the {\it volume element}.
\par
Generally speaking, in works about tensor algebra, the definition employed by Penrose and Rindler is used more frequently; in works utilizing the index-free notation of exterior forms, the ``geometric'' notation is preferred.
\par
How does this affect the wedge product? If the ``tensorial'' notation is employed, the wedge product of a $p$-form $A_{i_1\ldots i_p}$ and a $q$-form $B_{j_1\ldots j_q}$ is defined as:
\begin{equation}
A_{i_1\ldots i_p}\wedge B_{j_1\ldots j_q}=\frac{1}{(p+q)!}\delta_{i_1\ldots i_pj_1\ldots j_q}^{k_1\ldots k_pl_1\ldots l_q}A_{k_1\ldots k_p}B_{l_1\ldots l_q}
\label{tenwedge}
\end{equation}
In the ``geometric'' case, however, the definition is altered:
\begin{equation}
A_{i_1\ldots i_p}\wedge B_{j_1\ldots j_q}=\frac{1}{p!q!}\delta_{i_1\ldots i_pj_1\ldots j_q}^{k_1\ldots k_pl_1\ldots l_q}A_{k_1\ldots k_p}B_{l_1\ldots l_q}
\label{geowedge}
\end{equation}
\par
The exterior derivative is formally defined as the wedge product of the partial derivative operator and a tensor. Consequently, this definition is also dependent on how we define the wedge product. In the ``tensorial'' case, the exterior derivative is defined as
\begin{equation}
\mathrm{d}A=\nabla_{[i}A_{j_1\ldots j_p]}=\frac{1}{(p+1)!}\delta_{ij_1\ldots j_p}^{kl_1\ldots l_p}\nabla_kA_{l_1\ldots l_p}
\end{equation}
In the ``geometric'' interpretation, the definition of the exterior derivative changes to
\begin{equation}
\mathrm{d}A=(p+1)\nabla_{[i}A_{j_1\ldots j_p]}=\frac{1}{p!}\delta_{ij_1\ldots j_p}^{kl_1\ldots l_p}\nabla_kA_{l_1\ldots l_p}
\end{equation}
\par
Fully antisymmetric covariant tensors, like any other tensors, can be contracted with a vector: e.g., $v^iA_{i\ldots k}$. What makes the case of fully antisymmetric covariant tensors unique is that the index by which the contraction takes place does not matter: if a different index is used, only the sign of the result will be affected, and the sign change will be consistent with the permutation of indices.
\subsection{Abstract tensor algebras}
Tensors, and matrices representing tensors, offer rich algebraic structures. As a simple example, $2\times2$-matrices in the form {\tiny$\pmatrix{x&y\cr-y&x}$} can be used to represent the algebra of complex numbers. Other examples include the Pauli and Dirac matrices, or the algebra of quaternions.
\par
One of the simplest algebra types is the {\it symmetric algebra}, defined by the commutation rule $u\cdot v-v\cdot u=0$.
\par
{\it Grassmann algebras} are anticommutative: $u\cdot v+v\cdot u=0$. The direct sum of all exterior forms in dimension $n$ forms a Grassmann algebra.
\par
A {\it Clifford algebra} (see \cite{D1994}) is defined by the scalar unit, one or more basis vectors, and a symmetric scalar anticommutator: $u\cdot v+v\cdot u=2f_s(u,v)$. Perhaps the simplest example is $\rm{R}(0,1)$, a Clifford algebra with 1 basis vector ($v_1$), and an anticommutator function that is defined as $f_s(v_1,v_1)=-1$. This is none other than the algebra of complex numbers.
\par
Similarly, the Clifford algebra $\rm{R}(0,2)$ with a anticommutator defined as $f_s(v_1,v_1)=f_s(v_2,v_2)=-1$, and 0 for all other arguments, defines the algebra of quaternions; $v_1$, $v_2$, and their product, $v_1\cdot v_2$ will correspond with the three quaternionic imaginary units.
\par
Other, notable Clifford algebras include the algebra of Pauli spin matrices ($\rm{R}(3,0)$) and Dirac spin matrices ($\rm{R}(3,1)$).
\par
Closely related are the {\it symplectic algebras}: these are defined by an antisymmetric scalar commutator function, such that $u\cdot v-v\cdot u=2f_a(u,v)$.
\par
{\it Lie enveloping algebras} are characterized by the Jacobi-identity: $\left[u,\left[v,w\right]\right]+\left[v,\left[w,u\right]\right]+\left[w,\left[u,v\right]\right]=0$, where $\left[,\right]$ denotes the commutator, an antisymmetric vector-valued function: $\left[u,v\right]=u.v-v.u=2\vec{v}_a(u,v)$.
\section{Tensors in Maxima}
A computer algebra system can represent tensor quantities in many ways.
\par
First, tensors can be represented in terms of their components relative to some coordinate system. Second, tensors can be represented by their index properties. Third, tensors can be represented by symbols, with special simplification rules used to resolve expressions containing these symbols.
\par
All these representations are implemented by the various Maxima tensor packages: ATENSOR, CTENSOR, and ITENSOR.
\subsection{Indicial tensor manipulation}
The ITENSOR package treats tensors as ``opaque'' objects. Tensor values do not matter; tensor expressions are evaluated using formal algebraic rules, in particular algebraic rules concerning abstract tensor indices. In order to understand how this package works, it is nececssary first to take a brief look at how Maxima represents algebraic objects.
\subsubsection{The Maxima architecture}
It is not the purpose of this paper to provide a comprehensive overview of the Maxima system. (The author must admit that many areas of this large, complex system remain to him a complete mystery.) What is relevant with respect to the ITENSOR package is how Maxima represents expressions, function calls and index expressions in particular, using the LISP language.
\par
The core concept of the LISP language is, unsurprisingly, the idea of a {\it list}. Or, to be more precise, the idea of an ordered pair, the first element of which is the head (or {\tt car}), the second element the tail (or {\tt cdr}) of the list. List elements are themselves either lists or {\it atoms}: e.g., a number, a symbol, the empty list ({\tt nil}).
\par
For instance, the list {\tt (1 2 3)} would be represented as the ordered pair of the elements {\tt 1}, and {\tt (2 3)}. The first of these is an atom, the second itself a list, i.e., another ordered pair. The first element of this second pair is the atom {\tt 2}, the second the {\em list} {\tt (3)}. This last list, too, is represented by an ordered pair: the first element is the atom {\tt 3}, the second element the empty list {\tt nil}. In diagram form:
\par
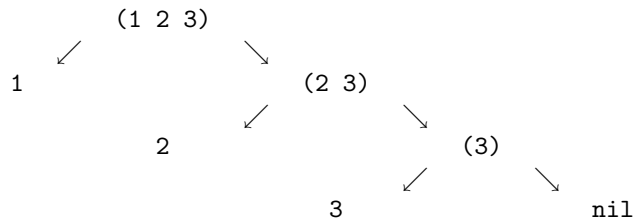
\begin{center}
\begin{tabular}{ccccccccc}
&&{\tt (1 2 3)}\\
&$\swarrow$&&$\searrow$\\
{\tt 1}&&&&{\tt (2 3)}\\
&&&$\swarrow$&&$\searrow$\\
&&{\tt 2}&&&&{\tt (3)}\\
&&&&&$\swarrow$&&$\searrow$\\
&&&&{\tt 3}&&&&{\tt nil}\\
\\
\end{tabular}
\end{center}
\par
LISP provides operator functions that are used to extract elements or sublists from a list. The most common of these are {\tt car} (to obtain the head of a list) and {\tt cdr} (to obtain the tail). In other words, {\tt car} gives you the first of the ordered pair of elements, {\tt cdr} the second.
\par
All LISP implementations provide many convenience functions that are combinations of {\tt car} and {\tt cdr}. For instance, {\tt cadr} is the head of the tail of a list: e.g., {\tt (cadr '(1 2 3))} gives the atom {\tt 2}. In other words, {\tt (cadr x)} is shorthand for {\tt (car (cdr x))}.
\par
LISP is a particularly suitable choice of a programming language for representing mathematical expressions. LISP has a strict ``prefix'' syntax: operators, be they unary, binary operators or function calls, always precede their list of arguments. As an example, here is a possible LISP representation of the mathematical expression $x+y\sin{z}$:
\par
\begin{verbatim}
(+ x (* y (sin z)))
\end{verbatim}
\par
Maxima uses similarly constructed LISP lists to represent expressions. However, instead of LISP operators, which may not always perform in the desired manner, Maxima implements its own set of operator functions. So the same mathematical expression looks slightly different as a Maxima object:
\par
\begin{verbatim}
((mplus) $x ((mtimes) $y (($sin) $z)))
\end{verbatim}
\par
The dollar sign in these expressions simply indicates a Maxima variable, as distinguished from LISP-level keywords or variables.
\par
As a computer algebra system, Maxima has a powerful simplification capability. When an expression is entered by the user, it is automatically processed by the simplifier. To avoid unnecessary processing, the simplifier ``marks'' expressions in various ways. After being processed by the simplifier, the previous expression will appear thus:
\par
\begin{verbatim}
((mplus simp) $x ((mtimes simp) $y (($sin simp) $z)))
\end{verbatim}
\par
Maxima also has lists. These are to be distinguished from LISP lists. In the syntax of the Maxima language, a Maxima list is a comma-separated list of objects enclosed in square brackets. Internally, the Maxima list {\tt[a,b,c]} is represented in LISP as:
\par
\begin{verbatim}
((mlist) $a $b $c)
\end{verbatim}
\par
\subsubsection{Indexed objects}
Tensors are physical/geometric quantities that are characterized by their transformation rules under coordinate transformations. Though it is often convenient to think of tensors as matrices, a matrix is nothing more but the representation of a tensor in a particular coordinate system; when the coordinate system changes, so do the elements of the matrix, even though the tensorial quantity they represent remains the same.
\par
Nevertheless, a number of tensorial operations exist that can be carried out in an arbitrary coordinate system, yet the result would be a proper representation of the appropriate tensorial quantity in that coordinate system. This means that a coordinate transformation can be applied either before or after the operation is performed, and the result would be the same. Symbolically, if $f$ is a tensor operator acting on a set of tensors $T_1,T_2\ldots T_n$, and $'$ denotes the coordinate transformation, the following is sometimes true:
\begin{equation}
\left[f(T_1,T_2\ldots T_n)\right]'=f(T'_1,T'_2\ldots T'_n)
\end{equation}
In particular, $f$ can be tensor addition, the inner or outer product of tensors, and certain types of tensor differentiation.
\par
This gives rise to the abstract index notation which one frequently encounters in the literature: we do not care what coordinate system is used, but we employ indices to unambiguously specify the operation.
\par
Abstract indices can be covariant and contravariant.
\par
One reason why the abstract index notation proves to be a very powerful algebraic tool is a ``feature'' of the inner product operation: index contraction. For instance, it is known that the metric tensor, $g$, contracts with other tensors, and can be used to raise or lower tensor indices. So for instance, the following is treated as an identity:
$$
g_{ab}T^{bc}=T_a^c
$$
\par
Notable is the fact that this operation can be carried out without knowing anything about what's ``inside'' the tensor $T$: the rules are strictly formal in nature, for manipulating the ``indexed object'' $T$, and the only information needed to carry out the operation is the pattern of indices that the object possesses.
\par
\subsubsection{Representing tensors in ITENSOR}
It is this formalism that is captured by the ITENSOR package. At the risk of sounding like a marketing brochure, it must be mentioned that it is this capability that makes ITENSOR particularly unique and powerful: it can be used for many tensorial problems very efficiently, as no computational resources are wasted to manage the ``contents'' of a tensor when, in fact, algebraic manipulation of the indexed symbols can suffice.
\par
The ITENSOR package represents tensors as formal Maxima function calls. For instance, the tensor $T_a^c$ would be represented as:
\par
\begin{verbatim}
T([a],[c])
\end{verbatim}
\par
Many of the features of ITENSOR can be discerned from this simple example. The tensor $T$ is formally represented as a function call with two parameters. Both parameters are Maxima lists: the first contains the covariant, the second, the contravariant indices of the tensor.
\par
The ITENSOR package includes a function whose purpose is to visually present such indexed objects in the familiar notation. This is illustrated by the following excerpt from a Maxima session:
\par
\begin{verbatim}
(%i2) ishow(g([a,b],[]))$
(%t2)                                g
                                      a b
\end{verbatim}
\par
An experienced user would frequently use the {\tt ishow} function inside tensorial expressions in order to present interim results in a visually pleasing form.
\par
As remarked earlier, a Maxima function call is represented internally as a LISP list. Accordingly, the tensor $g_{ab}$ is represented internally by ITENSOR as follows:
\par
\begin{verbatim}
((mlist) $g ((mlist) $a $b) ((mlist)))
\end{verbatim}
\par
Note that although the list of contravariant indices is empty, it is nevertheless present.
\par
Tensors in ITENSOR can also have derivative indices. These indices represent ordinary partial derivatives in the covariant position. For instance, the tensor $g_{ab,c}$ would be expressed using ITENSOR as {\tt g([a,b],[],c)}. (Note that the similar, widely used notation to represent covariant differentiation, e.g., $g_{ab;c}$, is not supported by the current version of ITENSOR.)
\par
The LISP list corresponding with the tensor $g_{ab,c}$ is
\par
\begin{verbatim}
(($g) ((mlist) $a $b) ((mlist)) $c)
\end{verbatim}
\par
This formalism makes it particularly easy to extract the list of covariant, contravariant, and derivative indices of a tensor. Once it has been ascertained that, say, {\tt \$x} is a tensor object, the LISP expressions {\tt (cdadr \$x)}, {\tt (cdaddr \$x)}, and {\tt (cdddr \$x)} can be used to obtain the list of covariant, contravariant, and derivative indices, respectively. 
\par
Tensor contraction is demonstrated by the following excerpt:
\par
\begin{verbatim}
(%i3) imetric(g);
(%o3)                                done
(%i4) ishow(g([b,c],[])*T([],[a,b]))$
                                    a b
(%t4)                              T    g
                                         b c
(%i5) ishow(contract(%))$
                                       a
(%t5)                                 T
                                       c
\end{verbatim}
\par
Incidentally, this example already demonstrates one of the shortcomings of the notation employed in ITENSOR. When the index just lowered is raised again, we should get back the original tensor: i.e., $g^{dc}g_{bc}T^{ab}=g^{dc}T^a{}_c=T^{ad}$. However, when we carry out the computation in Maxima, we get a different result:
\par
\begin{verbatim}
(%i6) ishow(contract(g([],[d,c])*%))$
                                      d a
(%t6)                                T
\end{verbatim}
\par
The reason for this discrepancy is due to the fact that ITENSOR maintains contravariant and covariant indices in two separate lists. When a contravariant index is lowered, information regarding its position relative to other contravariant indices is lost; when the index is subsequently raised, its position among contravariant indices will be predetermined by the contraction algorithm, and will not in any way be related to its location prior to its lowering.
\subsubsection{A new index notation}
This is one of the shortcomings in ITENSOR that was addressed by the present work. The goal was to develop a notation that is a) fully compatible with the current ITENSOR notation and can be used interchangeably with it; b) is relatively easy to implement, requiring only minor modifications to ITENSOR's complex algorithms; and c) provides a tensor notation that can be used to keep track of index ordering in arbitrarily complex tensor expressions.
\par
Sometimes, it is best not to reinvent the wheel. For this reason, the present author examined how this problem has been addressed in other tensor algebra packages. For instance, the built-in tensor algebra package in Maple 9.5 uses the relatively cumbersome method of explicitly spelling out the position of indices using the ``index characteristic'' component of a tensor object. Tensor objects are thus defined using expressions like this one:
\par
\begin{verbatim}
> g:=create([-1,-1],some_array);
\end{verbatim}
\par
Though unambiguous, this notation does not make it easy to enter and manipulate complex equations involving many indexed objects.
\par
The downloadable grTensorII package for Maple, designed specifically for use in general relativity, employs a similar notation.
\par
The Derive package, currently marketed by Texas Instruments, uses a somewhat more efficient representation. In Derive, the underscore character in the tensor's name is used to separate groups of covariant and contravariant indices. For instance, the covariant metric tensor, $g_{ij}$, in Derive is denoted as {\tt g\_ij}, while the (1,3) Riemann tensor $R^i{}_{jkl}$ would be denoted as {\tt R\_\_i\_jkl}.
\par
Both packages suffer from one serious shortcoming, however. They do not conceptually separate the components of a tensor from the algebraic symbol of the tensor. As has been remarked above, many tensorial equations can be computed utilizing only the algebraic properties of the tensors involved; carrying the components, in these cases, would mean an unnecessary computational burden.
\par
In any case, while each of these notational conventions has merits, they are not directly applicable to ITENSOR, since they substantially differ from the indexed object notation employed by this package. It is obvious that whatever convention is used, the basic formalism of a tensor as a formal function call with indices supplied in the form of list arguments should be preserved.
\par
As such, two possible extensions to the tensor syntax present themselves. One possibility, inspired by Derive, is to allow multiple groups of covariant and contravariant indices. For instance, the tensor $T_a{}^b{}_c{}^d$ could be represented as the Maxima object {\tt T([a],[b],[c],[d])} where each list in an odd position contains covariant indices, while lists in even positions contain contravariant indices. Index raising and lowering would move indices only to neighboring lists, which would ensure that an index that is raised and then subsequently lowered retains its original position relative to other indices.
\par
The main shortcoming of this notation is that it is still relatively cumbersome to use. It also makes it somewhat difficult to extract derivative indices; the algorithm would have to enumerate all parameters to the formal function call that represents a tensor, find the first parameter that is not a list, and extract this and the remaining parameters. It would also be quite difficult to adapt many of the utility functions of ITENSOR to successfully deal with this notation.
\par
For this reason, eventually another notation was chosen. To understand the rationale for this choice, first notice that in the abstract index formalism indices are truly abstract: they merely serve as labels, nothing more. In particular, no valid tensor equation contains mathematical expressions in the place of an index: $T_{i+1}$ is not a valid symbol.
\par
What this means is that we are free to use some mathematical operator to ``mark'' indices. Specifically, nothing prevents us from using the negative sign, for instance, to mark indices that are in the ``wrong'' list: e.g., contravariant indices that appear in the list of covariant indices, or vice versa.
\par
In other words, in this notation the tensor $T_a{}^b{}_c{}^d$ could be represented as the object {\tt T([a,-b,c,-d],[])}. This would be effectively equivalent to the ``old'' notation of {\tt T([a,c],[b,d])} with one crucial difference: the relative ordering between covariant and contravariant indices is preserved.
\par
How difficult is this notation to implement, while permitting it to coexist with the ``old'' notation? Fairly easy, as it turns out. For most ITENSOR functions, the actual index ordering doesn't matter, so in these cases, all we need is new functions to extract the list of covariant and contravariant indices. These tasks can be accomplished easily using two helper functions that extract the ``positive'' and ``negative'' indices from a list. In order to understand these helper functions, it is necessary to know how Maxima represents the negative sign: what is used is the expression $-1\cdot x$, as in {\tt ((mtimes) -1 \$x)}, to represent $-x$. Thus we have the following LISP code:
\par
\begin{verbatim}
(defun plusi(l)
  (cond
    ((null l) l)
    ((atom (car l))  (cons (car l) (plusi (cdr l))))
    ((and (eq (caaar l) 'mtimes) (eq (cadar l) -1)) (plusi (cdr l)))
    (t (cons (car l) (plusi (cdr l))))
  )
)
\end{verbatim}
\begin{verbatim}
(defun minusi(l)
  (cond
    ((null l) l)
    ((atom (car l))  (minusi (cdr l)))
    (
      (and (eq (caaar l) 'mtimes) (eq (cadar l) -1))
      (cons (caddar l) (minusi (cdr l)))
    )
    (t (minusi (cdr l)))
  )
)
\end{verbatim}
\par
With these two functions, obtaining the list of covariant and contravariant indices is extremely simple:
\par
\begin{verbatim}
(defun covi (rp) (plusi (cdadr rp)))
(defun conti (rp) (append (minusi (cdadr rp)) (cdaddr rp)))
\end{verbatim}
\par
Armed with these two functions, all that was left to be done is to substitute {\tt (covi x)} and {\tt (conti x)} in place of {\tt (cdadr x)} and {\tt (cdaddr x)}, respectively, in all the appropriate places. (A slight complication arises from the fact that in some cases, the ``pure'' list of indices was obtained, while in other cases, a Maxima-style list which begins with the object {\tt (mlist)} was needed, but this is an implementation detail.)
\par
The place where index order does matter is inside the {\tt contract} function. More specifically, inside the (internal) helper function {\tt contract1}. The following diagram depicts the various contraction functions that implement index contraction for different object types:
\par
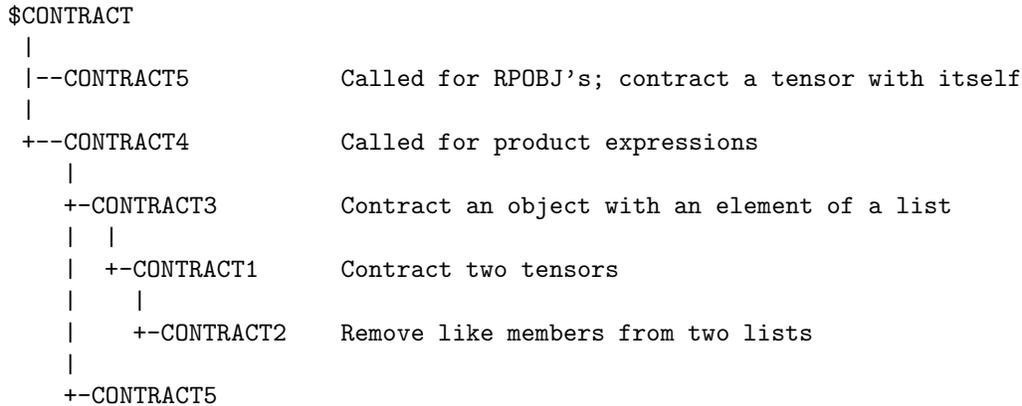
\begin{verbatim}
    $CONTRACT
     |
     |--CONTRACT5           Called for RPOBJ's; contract a tensor with itself
     |
     +--CONTRACT4           Called for product expressions
        |
        +-CONTRACT3         Contract an object with an element of a list
        |  |
        |  +-CONTRACT1      Contract two tensors
        |    |
        |    +-CONTRACT2    Remove like members from two lists
        |
        +-CONTRACT5
\end{verbatim}
As this diagram shows, the actual contraction of two tensors is carried out by {\tt contract1}. Here is the relevant fragment of this function that checks if there are covariant indices ({\tt c}) in an object that match the contravariant indices ({\tt b}) of the contracting object, or contravariant indices ({\tt d}) in an object that match the covariant indices ({\tt a}) in the contracting object:
\par
\begin{verbatim}
    (cond
      (
        (and b c (setq f (contract2 b c)))
        (setq b (car f) c (cdr f))
      )
      (
        (and a d (setq f (contract2 a d)))
        (setq a (car f) d (cdr f))
      )
\end{verbatim}
\par
(The function {\tt contract2} takes two lists of indices as its parameters and returns a pair of lists: the {\tt car} contains the ``free'' indices in the first argument, while the {\tt cdr} contains the ``free'' indices in the second argument.)
\par
This area of the code needed to be expanded, in order to accommodate the new style indices. Two separate cases had to be addressed: when the covariant list of the contracting object contained any ``negative'' indices that matched covariant indices in the other object, or if the covariant list of the contracting object contained any ``positive'' indices that appeared as ``negative'' indices in the covariant list of the other object. These two tasks are accomplished by the following code fragment:
\begin{verbatim}
    (cond
    ...
      (
        (and a (minusi c) (setq f (contract2 a (minusi c))))
        (do
          (
            (i c (cdr i))
            (j (car f))
            (k)
          )
          ((null i) (setq a (removenotin j a) c (reverse k)))
          (cond
            (
              (or (atom (car i)) (member (caddar i) (cdr f)))
              (setq k (cons (car i) k))
            )
            (
              (not (null j))
              (setq k (cons (car j) k) j (cdr j))
            )
          )
        )
      )
      (
        (and (minusi a) c (setq f (contract2 (minusi a) c)))
        (do
          (
            (i c (cdr i))
            (j (car f))
            (k)
          )
          ((null i)
            (setq
              c (reverse k)
              a (append
                (plusi a)
                (mapcar #'(lambda (x) (list '(mtimes simp) -1 x)) j)
              )
            )
          )
          (cond
            ((member (car i) (cdr f)) (setq k (cons (car i) k)))
            (
              (not (null j))
              (setq k (cons (list '(mtimes simp) -1 (car j)) k) j (cdr j))
            )
          )
        )
      )
\end{verbatim}
\par
(As its name suggests, the {\tt removenotin (i l)} helper function removes objects not in the list {\tt i} from the list {\tt l} and returns the truncated list as a result.)
\par
Surprisingly, one area of the ITENSOR package did not need to be modified: simplification functions that use the symmetry properties of tensorial objects to simplify tensor expressions.
\par
In ITENSOR, it is possible to declare a tensor to be symmetric in some of its indices using the {\tt decsym} function. For instance, the metric tensor is typically symmetric in both its covariant and in its contravariant form. Assuming that {\tt g} is the metric tensor, these symmetry properties would be expressed using the following commands:
\par
\begin{verbatim}
decsym(g,2,0,[sym(all)],[]);
decsym(g,0,2,[],[sym(all)]);
\end{verbatim}
\par
This notation remains fully applicable when ``new'' style tensor objects are used. In fact, the new notation makes things simpler for the Maxima user. The single declaration
\begin{verbatim}
decsym(g,2,0,[sym(all)],[]);
\end{verbatim}
is sufficient to express that $g_{ab}=g_{ba}$, $g^{ab}=g^{ba}$, and $g_a{}^b=g^b{}_a$, so long as all indices are placed in the first, ``covariant'' list, with a minus sign used to denote contravariant indices.
\subsubsection{Riemannian geometry}
The ITENSOR package contains built-in definitions for the Christoffel symbols of both kinds, and the (1,3) Riemann-tensor, expressed in terms of the metric tensor.
\par
With the help of the Christoffel-symbols, ITENSOR could express the covariant derivative of any tensor field, using the {\tt covdiff} function. This function, unfortunately, was broken in ITENSOR. The work to make ITENSOR fully functional again began here, by fixing this subroutine.
\par
The first new feature that was implemented in this area was the ability of ITENSOR to handle differentiated forms of the Christoffel-symbols. Previously, an expression in the form of, say, {\tt ichr2([a,b],[c],d)} resulted in an error. Now, however, these forms are correctly evaluated in terms of the metric tensor.
\par
A far more ambitious set of changes was implemented on the basis of the reference manual of the commercial version of MACSYMA (\cite{M1996}). The ITENSOR package in this product possessed the ability to perform computations not only in a coordinate base but also in a frame base; furthermore, it had the ability to deal with torsion and nonmetricity.
\par
A frame base is characterized by two entities: the basis vectors $e_{(a)}^i$, which can be collectively expressed in tensorial form, and the frame metric $\eta_{ab}$. The latter would normally be either the identity matrix (for a Euclidean orthonormal frame) or the Lorentz-matrix (for a frame with Minkowski signature.)
\par
Note that ITENSOR does not have the capability to distinguish frame and coordinate indices. This can lead to ambiguities in ITENSOR expressions. This represents few problems so long as 1) the same symbol is not reused as both a frame and a coordinate index, to avoid contraction between incompatible indices, and 2) differentiation is carried out on expressions containing either only coordinate or only frame indices, with the {\tt iframe\_flag} flag set appropriately. See the discussion at the end of section 2.3 of \cite{PR1990} for details.
\par
When a frame base, torsion, or nonmetricity is used, in most expressions involving curvature, the Christoffel-symbols are replaced by connection coefficients:
\par
\begin{equation}
c_{abc}=\Gamma_{abc}-\kappa_{abc}-\nu_{abc}
\label{cccoord}
\end{equation}
when a coordinate base used, or
\begin{equation}
c_{abc}=\gamma_{abc}-\nu_{abc}
\label{ccframe}
\end{equation}
in case of a frame base. Here, $\gamma$ are the frame coefficients (including torsion), the symbol $\kappa$ denotes the contortion coefficients, while $\nu$ denotes the nonmetricity coefficients.
\par
In the current version of ITENSOR, all these coefficients have now been implemented. While it is recognized not to be mathematically exhaustive, as a test the following relationships can be verified. First, when torsion is introduced:
\par
\begin{equation}
f_{;ij}-f_{;ji}=\left(\kappa_{ij}{}^k-\kappa_{ji}{}^k\right)f_{,k}=-\tau_{ij}^k f_{,k}
\end{equation}
if $\tau$ is antisymmetric in its covariant indices. For nonmetricity, the following applies:
\begin{equation}
g_{ij;k}=-\mu_kg_{ij}
\end{equation}
where $g$ denotes the metric tensor.
\subsubsection{Contraction of non-symmetric tensors}
Previously, when the ITENSOR package contracted tensor expressions, it did not take into account index ordering. Thus, for instance, $\epsilon^{ab}T_{bc}$ and $\epsilon^{ba}T_{bc}$ both yielded the same result, even if $\epsilon$ was declared as an antisymmetric tensor. The new implementation relies on the {\tt canform} function to determine the canonical ordering of indices and any resulting sign changes before executing the contraction, thereby ensuring that the contraction operation provides consistent results.
\subsubsection{Exterior algebra}
Thanks in a very large part to the work of Valery Pipin, ITENSOR now has capabilities relating to fully antisymmetric covariant tensors. Paralleling the functionality of Maxima's {\tt cartan} package (which introduces an index-free formalism for differential forms), ITENSOR can now compute wedge products, contraction of an exterior form with a vector, and the exterior derivative. Furthermore, ITENSOR also offers the capability to compute the Lie-derivative of a tensor of arbitrary valence.
\par
\par
Since ITENSOR is a tensor algebra package, it appears natural to use the ``tensorial'' definition (\ref{tenwedge}) of exterior forms. Many problems, however, can be expressed more naturally when the ``geometric'' definition (\ref{geowedge}) is employed. For this reason, a flag has been implemented in ITENSOR, which allows the user to use whichever definition is more applicable to the problem at hand. The {\tt igeowedge\_flag} variable is set to {\tt false} by default, but if it is set to {\tt true}, the ``geometric'' interpration is employed.
\subsubsection{Other changes}
One other major change that was implemented in the ITENSOR package was the consistent renaming of all objects and functions in order to conform to the naming conventions of commercial MACSYMA. The new names also make it a great deal less likely for ITENSOR names to conflict with names in the core Maxima system, or in other Maxima packages.
\subsection{Component tensor manipulation}
The component tensor manipulation package of Maxima, CTENSOR, views tensors as matrices. The main purpose of this package is to compute frequently used tensors in the coordinate system of the user's choice, given the metric that the user provides. As an example, CTENSOR can be used to compute the Ricci-tensor of a specific metric and show, after simplification, whether or not the metric is that of an empty spacetime.
\par
Until recently, the CTENSOR package was largely non-functional. If the user tried to follow simple examples and invoked the {\tt csetup} function, the program became apparently non-responsive, and Maxima had to be shut down. This problem turned out to be entirely superficial in nature: other components of Maxima failed to flush output buffers, causing an apparent lockup of the system when Maxima was waiting for user input, but the corresponding prompts were not yet displayed.
\par
Most of the work in CTENSOR since involved new features, most notably the support for frames, torsion, nonmetricity, and the algebraic classification of metrics.
\par
The CTENSOR package now has the capability to compute all the important tensors used in differential geometry. These include the Christoffel-symbols of the first and second kind, the Riemann tensor, the Ricci-tensor, the Einstein tensor, and the Weyl tensor. These tensors are calculated after the user has specified the coordinate variables, and entered the metric tensor in the form of a matrix, with elements that are functions of the coordinates. Special simplifications are applied when the metric tensor is diagonal, speeding up calculations considerably. These tools make many useful calculations possible; for instance, one demonstration file shows how Schwarzschild's spherically symmetric solution to Einstein's field equations can be derived.
\subsubsection{Using a frame base}
\par
A rigid frame is characterized by its basis vectors $e_{(a)}$, where labels in parentheses are basis vector labels, not coordinate indices. We can, of course, also express the coordinates of the basis vectors relative to some coordinate system: the components of these (contravariant) vectors will be $e_{(a)}^i$.
\par
\par
In its current version, the behavior of the CTENSOR package is now controlled by a flag, {\tt cframe\_flag}, which determines whether the Christoffel-symbols or the connection coefficients are used when computing curvature.
\par
When frames are used, the user specifies the metric by providing the covariant components of the frame base. (This is consistent with the convention of specifying a metric in a coordinate base by the covariant components of the metric tensor.) The user must also enter the frame metric. A function is provided that computes the contravariant frame base components, and also calculates the metric tensor.
\par
Note that no inverse functionality is provided: CTENSOR cannot compute a frame base if the metric is given. Though in principle it is possible to establish a set of equations that an orthonormal frame base must satisfy, and from these compute a frame base, this blind, ``brute force'' method rarely, if ever, results in a frame base that is of any use.
\par
\subsubsection{Algebraic classification of metrics}
New functions have been added to CTENSOR that can compute the covariant and contravariant components of the Newman-Penrose tetrad and, using this tetrad and the Weyl-tensor, compute the Petrov-classification of metrics. The Petrov algorithm is a direct implementation of the algorithm described in \cite{PSD2000}. This package has so far been tested with some simple cases, and it correctly obtained the Petrov type of the Schwarzschild, Kerr and anti-deSitter metrics, G\"odel's Universe, the Taub-Nut plane vacuum, and Allnutt's perfect fluid metric. Some of these results were cross-checked against results obtained from the dedicated computer algebra system {\tt SHEEP}, used for research in general relativity. That said, it should be emphasized that to date, the {\tt petrov()} function has only been tested in these simplest of cases and a comprehensive test suite that can exercise all branches in the decision tree used by this function remains to be written.
\par
The actual algorithm for the Petrov classification is implemented by the following subroutine, written in the Maxima language:
\par
\begin{verbatim}
petrov():=block
(
  [
    T:[   0,'N,'II,'III, 'D,'II,'II, 7,
        'II,'I, 'I,  11,'II, 13, 14,15,
         'N,'I, 'I,  19,'II, 21, 13,23,
       'III,19, 11,  27,  7, 23, 15,31],
    P:1
  ],

  if psi[4]#0 then P:P+1,
  if psi[3]#0 then P:P+2,
  if psi[2]#0 then P:P+4,
  if psi[1]#0 then P:P+8,
  if psi[0]#0 then P:P+16,

  if numberp(T[P]) and T[P]>0 then
  (
    if T[P]=7 then
      if ratsimp(psi[3]^2-3*psi[2]*psi[4])=0 then 'D
      else 'II
    else if T[P]=11 then
      if ratsimp(27*psi[4]^2*psi[1]+64*psi[3]^3)=0 then 'II
      else 'I
    else if T[P]=13 then
      if ratsimp(psi[1]^2*psi[4]+2*psi[2]^3)=0 then 'II
      else 'I
    else if T[P]=14 then
      if ratsimp(9*psi[2]^2-16*psi[1]*psi[3])=0 then 'II
      else 'I
    else if T[P]=15 then
      if ratsimp(3*psi[2]^2-4*psi[1]*psi[3])=0 and
         ratsimp(psi[2]*psi[3]-3*psi[1]*psi[4])=0 then 'II
      else 'I
    else if T[P]=19 then
      if ratsimp(psi[0]*psi[4]^3-27*psi[3]^4)=0 then 'II
      else 'I
    else if T[P]=21 then
      if ratsimp(9*psi[2]^2-psi[4]^2) then 'D
      else 'I
    else if T[P]=23 then block
    (
      [I:ratsimp(psi[0]*psi[4]+3*psi[2]^2)],
      if I=0 and ratsimp(4*psi[2]*psi[4]-3*psi[3]^2)=0 then 'III
      else block
      (
        [J:ratsimp(4*psi[2]*psi[4]-3*psi[3]^2)],
        if ratsimp(psi[4]*I^2-3*J*(psi[0]*J-2*psi[2]*I))=0 then 'II
        else 'I
      )
    )
    else if T[P]=27 then
      if ratsimp(psi[0]*psi[3]^2-psi[1]^2*psi[4])=0 then
        if ratsimp(psi[0]*psi[4]+2*psi[1]*psi[3])=0 then 'D
        else if ratsimp(psi[0]*psi[4]-16*psi[1]*psi[3])=0 then 'II
        else 'I
      else block
      (
        [I:ratsimp(psi[0]*psi[4]+2*psi[1]*psi[3])],
        if I=0 then block
        (
          [J:ratsimp(-psi[0]*psi[3]^2-psi[1]^2*psi[4])],
          if J=0 then 'III
          else if ratsimp(I^3-27*J^2)=0 then 'II
          else 'I
        )
        else 'I
      )
    else block
    (
      [H:ratsimp(psi[0]*psi[2]-psi[1]^2)],
      if H=0 then
        if ratsimp(psi[0]*psi[3]-psi[1]*psi[2])=0 then
          if ratsimp(psi[0]*psi[4]-psi[2]^2)=0 then 'N
          else 'I
        else block
        (
          [E:ratsimp(psi[0]*psi[4]-psi[2]^2)],
          if E=0 then
            if ratsimp(37*psi[2]^2+27*psi[1]*psi[3])=0 then 'II
            else 'I
          else block
          (
            [
              A:ratsimp(psi[1]*psi[3]+psi[2]^2),
              I:ratsimp(E-4*A)
            ],
            if I#0 and ratsimp(I^3-27*(psi[4]*H-psi[3]^2*psi[0]+
                               psi[1]*psi[2]*psi[3]+psi[2]*A)^2)=0 then 'II
            else 'I
          )
        )
        else block
        (
          [I:ratsimp(psi[0]*psi[4]-psi[2]^2-4*(psi[1]*psi[3]+psi[2]^2))],
          if I=0 then
            if ratsimp(psi[4]*H-psi[3]^2*psi[0]+psi[1]*psi[2]*psi[3]+
                       psi[2]*(psi[1]*psi[3]+psi[2]^2))=0 then 'III
            else 'I
          else
            if ratsimp(psi[0]^2*psi[3]-psi[0]*psi[1]*psi[2]-2*psi[1]*H)=0 then
              if ratsimp(psi[0]^2*I-12*H^2)=0 then 'D
              else if ratsimp(psi[0]^2*I-3*H^2)=0 then 'II
              else 'I
            else block
            (
              [J:ratsimp(psi[4]*H-psi[3]^2*psi[0]+psi[1]*psi[2]*psi[3]+
                         psi[2]*(psi[1]*psi[3]+psi[2]^2))],
              if J#0 and ratsimp(I^3-27*J^2)=0 then 'II
              else 'I
            )
        )
    )
  )
  else T[P]
);
\end{verbatim}
\par
For the simpler cases, the metric type is obtained directly from the lookup table {\tt T}. For the more complicated cases, additional coefficients are computed on an ``as and when needed'' basis, in order to minimize the amount of computation required.
\subsubsection{Torsion and nonmetricity}
Torsion and nonmetricity can be represented as additional components in the connection coefficients. As defined in (\ref{cccoord}) and (\ref{ccframe}), the connection coefficients can incorporate torsion and nonmetricity using the contortion and nonmetricity coefficients. Two functions have been added to the CTENSOR package to compute these coefficients from a user-supplied torsion tensor and nonmetricity vector.
\subsubsection{Predefined metrics}
Another useful feature found in the commercial MACSYMA version of CTENSOR was predefined metrics. Through the {\tt ct\_coordsys()} function, CTENSOR had the ability to set up a variety of commonly used metrics in a coordinate or frame base.
\par
This capability has also been reproduced in the current version of CTENSOR. Specifically, the following metrics have been implemented:
\par
{\small
\begin{longtable}{cc}
Name/Coordinates&Metric ($g_{ik}$)/Frame ($e_{(a)i}$)\\
\hline\\
\begin{tabular}{c}cartesian2d\cr
$x,y$
\end{tabular}
&$\pmatrix{1&0\cr0&1}$, $\pmatrix{1&0\cr0&1}$\\\\
\hline\\
\begin{tabular}{c}polar\cr
$r,\phi$
\end{tabular}
&$\pmatrix{1&0\cr0&r^2}$, $\pmatrix{\cos{\phi}&-r\sin{\phi}\cr\sin{\phi}&r\cos{\phi}}$\\\\
\hline\\
\begin{tabular}{c}elliptic\cr
$u,v$\cr
$e$ is constant
\end{tabular}
&$\matrix{\pmatrix{e^2(\cosh^2{u}-\cos^2{v})&0\cr0&e^2(\cosh^2{u}-\cos^2{v})}\cr\\\cr\pmatrix{e\sinh{u}\cos{v}&-e\cosh{u}\sin{v}\cr e\cosh{u}\sin{v}&e\sinh{u}\cos{v}}}$\\\\
\hline\\
\begin{tabular}{c}confocalelliptic\cr
$u,v$\cr
$e$ is constant
\end{tabular}
&$\matrix{\pmatrix{\frac{e^2(u^2-v^2)}{u^2-1}&0\cr0&\frac{e^2(v^2-u^2)}{v^2-1}}\cr\\\cr\pmatrix{ev&eu\cr \frac{eu(1-v^2)}{\sqrt{(u^2-1)(1-v^2)}}&\frac{ev(1-u^2)}{\sqrt{(u^2-1)(1-v^2)}}}}$\\\\
\hline\\
\begin{tabular}{c}bipolar\cr
$u,v$\cr
$e$ is constant
\end{tabular}
&$\matrix{\pmatrix{\frac{e^2}{(\cosh{v}-\cos{u})^2}&0\cr0&\frac{e^2}{(\cosh{v}-\cos{u})^2}}\cr\\\cr\pmatrix{\frac{-e\sin{u}\sinh{v}}{(\cosh{v}-\cos{u})^2}&\frac{e(1-\cos{u}\cosh{v})}{(\cosh{v}-\cos{u})^2}\cr\frac{e(\cos{u}\cosh{v}-1)}{(\cosh{v}-\cos{u})^2}&\frac{-e\sin{u}\sinh{v}}{(\cosh{v}-\cos{u})^2}}}$\\\\
\hline\\
\begin{tabular}{c}parabolic\cr
$u,v$
\end{tabular}
&$\pmatrix{u^2+v^2&0\cr0&u^2+v^2}$, $\pmatrix{u&-v\cr v&u}$\\\\
\hline\\
\begin{tabular}{c}cartesian3d\cr
$x,y,z$
\end{tabular}
&$\pmatrix{1&0&0\cr0&1&0\cr0&0&1}$, $\pmatrix{1&0&0\cr0&1&0\cr0&0&1}$\\\\
\hline\\
\begin{tabular}{c}polarcylindrical\cr
$r,\theta,z$
\end{tabular}
&$\pmatrix{1&0&0\cr0&r^2&0\cr0&0&1}$, $\pmatrix{\cos{\theta}&-r\sin{\theta}&0\cr\sin{\theta}&r\cos{\theta}&0\cr0&0&1}$\\\\
\hline\\
\begin{tabular}{c}ellipticcylindrical\cr
$u,v,z$\cr
$e$ is constant
\end{tabular}
&$\matrix{\pmatrix{e^2(\sin^2{v}+\sinh^2{u})&0&0\cr0&e^2(\sin^2{v}+\sinh^2{u})&0\cr0&0&1}\cr\\\cr\pmatrix{e\sinh{u}\cos{v}&-e\cosh{u}\sin{v}&0\cr e\cosh{u}\sin{v}&e\sinh{u}\cos{v}&0\cr0&0&1}}$\\\\
\hline\\
\begin{tabular}{c}confocalellipsoidal\cr
$u,v,w$\cr
$e$, $f$, $g$ are constants
\end{tabular}
&$\matrix{\pmatrix{\frac{(v-u)(w-u)}{4(e^2-u)(u-f^2)(u-g^2)}&0&0\cr0&\frac{(v-u)(w-v)}{4(v-e^2)(v-f^2)(v-g^2)}&0\cr0&0&\frac{(w-u)(w-v)}{4(e^2-w)(w-f^2)(w-g^2)}}\cr\\\cr\pmatrix{
-\sqrt{\frac{(v-e^2)(w-e^2)}{4(f^2-e^2)(g^2-e^2)(e^2-u)}}&
-\sqrt{\frac{(u-e^2)(w-e^2)}{4(f^2-e^2)(g^2-e^2)(e^2-v)}}&
-\sqrt{\frac{(u-e^2)(v-e^2)}{4(f^2-e^2)(g^2-e^2)(e^2-w)}}\cr
 \sqrt{\frac{(v-f^2)(w-f^2)}{4(f^2-e^2)(g^2-f^2)(u-f^2)}}&
 \sqrt{\frac{(u-f^2)(w-f^2)}{4(f^2-e^2)(g^2-f^2)(v-f^2)}}&
 \sqrt{\frac{(u-f^2)(v-f^2)}{4(f^2-e^2)(g^2-f^2)(w-f^2)}}\cr
-\sqrt{\frac{(v-g^2)(w-g^2)}{4(g^2-e^2)(g^2-f^2)(g^2-u)}}&
-\sqrt{\frac{(u-g^2)(w-g^2)}{4(g^2-e^2)(g^2-f^2)(g^2-v)}}&
-\sqrt{\frac{(u-g^2)(v-g^2)}{4(g^2-e^2)(g^2-f^2)(g^2-w)}}
}}$\\\\
\hline\\
\begin{tabular}{c}bipolarcylindrical\cr
$u,v,z$\cr
$e$ is constant
\end{tabular}
&$\matrix{\pmatrix{\frac{e^2}{(\cosh{v}-\cos{u})^2}&\frac{2e^2\sin{u}\sinh{v}(1-\cos{u}\cosh{v})}{(\cosh{v}-\cos{u})^2}&0\cr\frac{2e^2\sin{u}\sinh{v}(1-\cos{u}\cosh{v})}{(\cosh{v}-\cos{u})^2}&\frac{e^2}{(\cosh{v}-\cos{u})^2}&0\cr0&0&\frac{e^2}{(\cosh{v}-\cos{u})^4}}\cr\\\cr\pmatrix{\frac{e\sin{u}\sinh{v}}{(\cosh{v}-\cos{u})^2}&\frac{e(1-\cos{u}\cosh{v})}{(\cosh{v}-\cos{u})^2}&0\cr\frac{e(\cos{u}\cosh{v}-1)}{(\cosh{v}-\cos{u})^2}&\frac{-e\sin{u}\sinh{v}}{(\cosh{v}-\cos{u})^2}&0\cr0&0&\frac{e}{(\cosh{v}-\cos{u})^2}}}$\\\\
\hline\\
\begin{tabular}{c}paraboliccylindrical\cr
$u,v,z$
\end{tabular}
&$\pmatrix{u^2+v^2&0&0\cr0&u^2+v^2&0\cr0&0&1}$, $\pmatrix{u&-v&0\cr v&u&0\cr0&0&1}$\\\\
\hline\\
\begin{tabular}{c}paraboloidal\cr
$u,v,\phi$
\end{tabular}
&$\pmatrix{u^2v^2&0&0\cr0&u^2+v^2&0\cr0&0&u^2+v^2}$, $\pmatrix{-uv\sin{\phi}&v\cos{\phi}&u\cos{\phi}\cr uv\cos{\phi}&v\sin{\phi}&u\sin{\phi}\cr0&u&-v}$\\\\
\hline\\
\begin{tabular}{c}conical\cr
$u,v,w$\cr
$e$, $f$ are constants
\end{tabular}
&$\matrix{\pmatrix{\frac{(v^2-u^2)w^2}{(u^2-e^2)(u^2-f^2)}&0&0\cr0&\frac{(u^2-v^2)w^2}{(v^2-e^2)(v^2-f^2)}&0\cr0&0&1}\cr\\\cr\pmatrix{\frac{vw}{ef}&\frac{uw}{ef}&\frac{uv}{ef}\cr\frac{-uw\sqrt{e^2-v^2}}{e\sqrt{(f^2-e^2)(u^2-e^2)}}&\frac{-vw\sqrt{e^2-u^2}}{e\sqrt{(f^2-e^2)(v^2-e^2)}}&\frac{\sqrt{(u^2-e^2)(v^2-e^2)}}{e\sqrt{e^2-f^2}}\cr\frac{uw\sqrt{v^2-f^2}}{e\sqrt{(f^2-e^2)(u^2-f^2)}}&\frac{vw\sqrt{u^2-f^2}}{f\sqrt{(f^2-e^2)(v^2-f^2)}}&\frac{\sqrt{(u^2-f^2)(v^2-f^2)}}{f\sqrt{f^2-e^2}}}}$\\\\
\hline\\
\begin{tabular}{c}toroidal\cr
$u,v,\phi$\cr
$e$ is constant
\end{tabular}
&$\matrix{\pmatrix{\frac{e^2\sinh^2{v}}{(\cosh{v}-\cos{u})^2}&0&0\cr0&\frac{e^2}{(\cosh{v}-\cos{u})^2}&0\cr0&0&\frac{e^2}{(\cosh{v}-\cos{u})^2}}\cr\\\cr\pmatrix{\frac{-e\sin{\phi}\sinh{v}}{(\cosh{v}-\cos{u})}&\frac{-e\cos{\phi}\sin{u}\sinh{v}}{(\cosh{v}-\cos{u})^2}&\frac{-e\cos{\phi}(\cos{u}\cosh{v}-1)}{(\cosh{v}-\cos{u})^2}\cr\frac{e\cos{\phi}\sinh{v}}{(\cosh{v}-\cos{u})}&\frac{-e\sin{\phi}\sin{u}\sinh(v)}{(\cosh{v}-\cos{u})^2}&\frac{-e\sin{\phi}(\cos{u}\cosh{v}-1)}{(\cosh{v}-\cos{u})^2}\cr0&\frac{e(\cos{u}\cosh{v}-1)}{(\cosh{v}-\cos{u})^2}&\frac{-e\sin{u}\sinh{v}}{(\cosh{v}-\cos{u})^2}}}$\\\\
\hline\\
\begin{tabular}{c}spherical\cr
$r,\theta,\phi$
\end{tabular}
&$\pmatrix{1&0&0\cr0&r^2&0\cr0&0&r^2\sin^2{\theta}}$, $\pmatrix{1&0&0\cr0&r&0\cr0&0&r\sin{\theta}}$\\\\
\hline\\
\begin{tabular}{c}oblatespheroidal\cr
$u,v,\phi$\cr
$e$ is constant
\end{tabular}
&$\matrix{\pmatrix{e^2(\sin^2{v}+\sinh^2{u})&0&0\cr0&e^2(\sin^2{v}+\sinh^2{u})&0\cr0&0&e^2\cosh^2{u}\cos^2{v}}\cr\pmatrix{|e|\sqrt{\sin^2{v}+\sinh^2{u}}&0&0\cr0&|e|\sqrt{\sin^2{v}+\sinh^2{u}}&0\cr0&0&|e|\cosh{u}|\cos{v}|}}$\\\\
\hline\\
\begin{tabular}{c}oblatespheroidalsqrt\cr
$u,v,\phi$\cr
$e$ is constant
\end{tabular}
&$\pmatrix{\frac{e^2(u^2-v^2)}{u^2-1}&0&0\cr0&\frac{e^2(u^2-v^2)}{u^2-1}&0\cr0&0&e^2u^2v^2}$, $\pmatrix{\frac{|e|\sqrt{u^2-v^2}}{\sqrt{u^2-1}}&0&0\cr0&\frac{|e|\sqrt{u^2-v^2}}{\sqrt{u^2-1}}&0\cr0&0&|euv|}$\\\\
\hline\\
\begin{tabular}{c}prolatespheroidal\cr
$u,v,\phi$\cr
$e$ is constant
\end{tabular}
&$\matrix{\pmatrix{e^2(\sin^2{v}+\sinh^2{u})&0&0\cr0&e^2(\sin^2{v}+\sinh^2{u})&0\cr0&0&e^2\sin^2{v}\sinh^2{u}}\cr\\\cr\pmatrix{|e|\sqrt{\sin^2{v}+\sinh^2{u}}&0&0\cr0&|e|\sqrt{\sin^2{v}+\sinh^2{u}}&0\cr0&0&|e\sinh{u}\sin{v}|}}$\\\\
\hline\\
\begin{tabular}{c}prolatespheroidalsqrt\cr
$u,v,\phi$\cr
$e$ is constant
\end{tabular}
&$\matrix{\pmatrix{\frac{e^2(v^2-u^2)}{1-u^2}&0&0\cr0&\frac{e^2(v^2-u^2)}{v^2-1}&0\cr0&0&e^2(1-u^2)(v^2-1)}\cr\\\cr\pmatrix{\frac{|e|\sqrt{v^2-u^2}}{\sqrt{1-u^2}}&0&0\cr0&\frac{|e|\sqrt{v^2-u^2}}{\sqrt{v^2-1}}&0\cr0&0&|e|\sqrt{(1-u^2)(v^2-1)}}}$\\\\
\hline\\
\begin{tabular}{c}ellipsoidal\cr
$r,\theta,\phi$\cr
$a$, $b$, $c$ are constants
\end{tabular}
&$\pmatrix{
(a^2\cos^2{\phi}+b^2\sin^2{\phi})\sin^2{\theta}+&
(a^2\cos^2{\phi}+b^2\sin^2{\phi}-c^2)\cdot&
(b^2-a^2)\cos{\phi}\sin{\phi}\cdot\\
+c^2\cos^2{\theta}&\cdot r\cos{\theta}\sin{\theta}&\cdot r\sin^2{\theta}\cr
(a^2\cos^2{\phi}+b^2\sin^2{\phi}-c^2)\cdot&
r^2((a^2\cos^2{\phi}+b^2\sin^2{\phi})\cos^2{\theta}+&
(b^2-a^2)\cos{\phi}\sin{\phi}\cdot\\
\cdot r\cos{\theta}\sin{\theta}&+c^2\sin^2{\theta})&\cdot r^2\cos{\theta}\sin{\theta}\cr
(b^2-a^2)\cos{\phi}\sin{\phi}\cdot&
(b^2-a^2)\cos{\phi}\sin{\phi}\cdot&
(a^2\sin^2{\phi}+b^2\cos^2{\phi})\cdot\\
\cdot r\sin^2{\theta}&\cdot r^2\cos{\theta}\sin{\theta}&\cdot r^2\sin^2{\theta}
}$\\\\&no frames\\\\
\hline\\
\begin{tabular}{c}cartesian4d\cr
$x,y,z,t$
\end{tabular}
&$\pmatrix{1&0&0&0\cr0&1&0&0\cr0&0&1&0\cr0&0&0&1}$, $\pmatrix{1&0&0&0\cr0&1&0&0\cr0&0&1&0\cr0&0&0&1}$\\\\
\hline\\
\begin{tabular}{c}spherical4d\cr
$r,\theta,\eta,\phi$
\end{tabular}
&$\pmatrix{1&0&0&0\cr0&r^2&0&0\cr0&0&r^2\sin^2{\theta}&0\cr0&0&0&r^2\sin^2{\eta}\sin^2{\theta}}$, $\pmatrix{1&0&0&0\cr0&r&0&0\cr0&0&r\sin{\theta}&0\cr0&0&0&r\sin{\eta}\sin{\theta}}$\\\\
\hline\\
\begin{tabular}{c}exteriorschwarzschild\cr
$t,r,\theta,\phi$\cr
$m$ is constant\cr
$r>2m$
\end{tabular}
&$\pmatrix{\frac{2m-r}{r}&0&0&0\cr0&\frac{r}{r-2m}&0&0\cr0&0&r^2&0\cr0&0&0&r^2\sin^2{\theta}}$, $\pmatrix{\sqrt{\frac{r-2m}{r}}&0&0&0\cr0&\sqrt{\frac{r}{r-2m}}&0&0\cr0&0&r&0\cr0&0&0&r\sin{\theta}}$\\\\
\hline\\
\begin{tabular}{c}interiorschwarzschild\cr
$t,z,u,v$\cr
$m$ is constant\cr
$t<2m$
\end{tabular}
&$\pmatrix{-\frac{t}{2m-t}&0&0&0\cr0&\frac{2m-t}{t}&0&0\cr0&0&t^2&0\cr0&0&0&t^2\sin^2{u}}$, $\pmatrix{\sqrt{\frac{t}{2m-t}}&0&0&0\cr0&\sqrt{\frac{2m-t}{t}}&0&0\cr0&0&t&0\cr0&0&0&t\sin{u}}$\\\\
\hline\\
\begin{tabular}{c}kerr\_newman\cr
$t,r,\theta,\phi$\cr
$a$, $m$ are constants
\end{tabular}
&$\matrix{\pmatrix{
\frac{2mr-r^2-a^2\cos^2{\theta}}{r^2+a^2\cos^2{\theta}}&0&0&\frac{-2amr\sin^2{\theta}}{r^2+a^2\cos^2{\theta}}\cr
0&\frac{r^2+a^2\cos^2{\theta}}{a^2-2mr+r^2}&0&0\cr
0&0&r^2+a^2\cos^2{\theta}&0\cr
\frac{-2amr\sin^2{\theta}}{r^2+a^2\cos^2{\theta}}&0&0&\frac{[r^4+2a^2r^2+a^4\cos^2{\theta}+(2a^2mr-a^2r^2)\sin^2{\theta}]\sin^2{\theta}}{r^2+a^2\cos^2{\theta}}
}\cr\\\cr\pmatrix{
\frac{\sqrt{a^2-2mr+r^2}}{\sqrt{r^2+a^2\cos^2{\theta}}}&0&0\frac{-a\sin^2{\theta}\sqrt{a^2-2mr+r^2}}{\sqrt{r^2+a^2\cos^2{\theta}}}\cr
0&\frac{\sqrt{r^2+a^2\cos^2{\theta}}}{\sqrt{a^2-2mr+r^2}}&0&0\cr
0&0&\sqrt{r^2+a^2\cos^2{\theta}}&0\cr
\frac{-a\sin{\theta}}{\sqrt{r^2+a^2\cos^2{\theta}}}&0&0&\frac{(r^2+a^2)\sin{\theta}}{\sqrt{r^2+a^2\cos^2{\theta}}}
}}$\\\\
\end{longtable}
}
\par
The signature of all these metrics is $(+,...,+)$ with the exception of the Schwarzschild and Kerr-Newman metrics that have a $(-,+,...+)$ signature. The {\tt ct\_coordsys} function also has the ability to add additional ``flat'' dimensions with either a Minkowski or a Euclidean signature.
\subsection{Algebraic tensor manipulation}
In addition to CTENSOR and ITENSOR, the commercial MACSYMA system included a third tensor package, ATENSOR. The purpose of this package was to capture the algebraic properties of ``general tensor algebras''. This package allowed the user to define a tensor algebra by specifying the algebra's commutation properties. A variety of algebras could be explored this way, including Clifford algebras, Grassmann algebras, and Lie algebras.
\par
To make a long story short, unlike CTENSOR and ITENSOR which, while in various states of disrepair, were nevertheless present in GPL Maxima, the ATENSOR package was completely missing. This has now been corrected, and an initial implementation of ATENSOR, fully conformant to the documentation of commercial MACSYMA, is now included with GPL Maxima.
\subsubsection{Commutators and abstract algebra types}
The ATENSOR package defines different algebra types based on their (anti)commutators. Several algebra types are recognized:
\par
\begin{tabular}{lc}
\\
Algebra type&Commutator\\
\hline\\[-6pt]
Universal&No commutation rules\\
Grassmann&$u\cdot v+v\cdot u=0$\\
Clifford&$u\cdot v+v\cdot u=2f_s(u,v)$\\
Symmetric&$u\cdot v-v\cdot u=0$\\
Symplectic&$u\cdot v-v\cdot u=2f_a(u,v)$\\
Lie enveloping&$u\cdot v-v\cdot u=2\vec{v}_a(u,v)$\\
\\
\end{tabular}
\par
Here, $f_s$ stands for a scalar valued symmetric function (i.e., $f_s(u,v)=f_s(v,u)$), $f_a$ stands for an antisymmetric scalar valued function ($f_a(u,v)=-f_a(v,u)$) and $\vec{v}_a$ is an antisymmetric vector valued function.
\par
The actual values of the functions $f_s$, $f_a$, and $\vec{v}_a$ are determined by the algebra's type and dimensionality. The ATENSOR package uses the matrix {\tt aform} in which the function values for base vector arguments are stored. For the vector-valued function $\vec{v}_a$, the matrix values are taken to be base vector indices. When the algebra type is selected by the user, this matrix is preinitialized appropriately.
\par
For this purpose, when a Clifford, Symplectic, or Lie enveloping algebra is selected, the user can optionally enter values that determine the algebra's dimensionality. For a Clifford algebra, up to three values can be used, specifying the positive, degenerate, and negative dimensions of the algebra, respectively. For a symplectic algebra, the number of regular and degenerate dimensions can be specified; for Lie enveloping algebras, a single number, the algebra's dimensionality, can be entered.
\par
For a Clifford algebra, the matrix {\tt aform} is initialized as ${\rm diag}(1\ldots 1,0\ldots 0,-1\ldots-1)$, where the number of 1's, 0's, and $-1$'s correspond with the number of positive, degenerate, and negative dimensions, respectively.
\par
For a symplectic algebra of (regular) dimension $n$, {\tt aform} is initialized to an $n\times n$ matrix with its off-diagonal values set to 1 or $-1$ depending on whether they represent an odd or even index permutation. If the algebra also has degenerate dimenions, the appropriate number of null columns and rows are added.
\par
Lastly, for Lie enveloping algebras {\tt aform} is initialized to an antisymmetric matrix whose elements are defined as $a_{ij}=\left({\rm mod}_n(2n+2-i-j)+1\right){\rm perm}(i,j)$
where ${\rm perm}(i,j)$ is a permutator function that gives $+1$ if $(i,j,1\ldots i-1,i+1,\ldots j-1,j+1,\ldots n)$ is an even permutation of the sequence $(1\ldots n)$, and $-1$ otherwise. For instance, the initialization function call {\tt init\_atensor(lie\_envelop,3)} produces the following matrix:
$$
\pmatrix{0&3&-2\cr-3&0&1\cr2&-1&0}
$$
\par
It is known (see, for instance, \cite{D1994}) that the Clifford algebra of 0 positive and $-2$ negative dimensions corresponds with the algebra of quaternions. The ATENSOR package can correctly reproduce the quaternionic multiplication table as
$$
\pmatrix{1&v_1&v_2&v_1\cdot v_2\cr v_1&-1&v_1\cdot v_2&-v_2\cr v_2&-v_1\cdot v_2&-1&v_1\cr v_1\cdot v_2&v_2&-v_1&-1}
$$
where the base vectors $v_1$, $v_2$, and their product $v_1\cdot v_2$ serve as the three quaternionic imaginary units. This was one of the tests used to ascertain that the ATENSOR package produces mathematically correct results.
\section{Conclusions}
The Maxima tensor packages remain a work in progress. It is the author's desire to maintain these packages in good working order so long as GPL Maxima itself remains actively maintained. It is hoped that the tensor packages will again be accepted by a user community, and as they are actively ``field tested'', it will be possible to make them more robust and mathematically accurate.
\bibliography{../refs}
\bibliographystyle{plain}
\end{document}